\documentclass[aps,prd,twocolumn,superscriptaddress,showpacs]{revtex4}
\usepackage{graphicx}
\usepackage{dcolumn}
\usepackage{bm}
\usepackage{natbib}
\usepackage{multirow}
\newcommand{\beq}{\begin{equation}}
\newcommand{\eeq}{\end{equation}}
\newcommand{\beqa}{\begin{eqnarray}}
\newcommand{\eeqa}{\end{eqnarray}}

\newcommand{\hGpc}{{\rm h^{-1}Gpc}}
\newcommand{\hMsun}{h^{-1}M_{\odot}}

\def\affilmrk#1{$^{#1}$}
\def\affilmk#1#2{$^{#1}$#2;}

\def\zrh{1}
\def\bkl{2}
\def\ewh{3}

\begin{document}

\title{How to suppress the shot noise in galaxy surveys}  
\author{
Uro\v s Seljak \affilmrk{\zrh,\bkl,\ewh}, Nico Hamaus \affilmrk{\zrh}, Vincent Desjacques 
}
\address{
{Institute for Theoretical Physics, University of Zurich, Zurich, Switzerland}
\affilmk{\bkl}{Physics Department and Lawrence Berkeley National Laboratory, University of California, Berkeley, California 94720, USA}
\affilmk{\ewh}{Ewha University, Seoul 120-750, S. Korea}
}
\date{\today}

\begin{abstract}

Galaxy surveys are one of the most powerful means to extract the cosmological information and for a given volume 
the attainable precision is determined by the galaxy shot noise $\sigma_n^2$ relative to the power spectrum $P$. 
It is generally assumed that shot noise is white and given by the inverse of the number density $\bar{n}$. 
In this paper we argue one may be able to considerably improve upon this: 
in the halo picture of cosmological structure all of the dark matter 
is in halos of varying mass and galaxies are formed inside these halos, but 
for the dark matter mass and momentum conservation guarantee that nonlinear effects cannot develop a white noise in the dark matter power spectrum on large scales. 
This suggests that with a suitable weighting a similar effect may be achieved for galaxies, suppressing their shot 
noise. We explore this idea with N-body simulations 
by weighting central halo galaxies by halo mass and find that the resulting shot noise 
can be reduced dramatically relative to expectations, with a 10-30 suppression 
at the highest number density of $\bar{n}=4\times 10^{-3}({\rm Mpc/h})^3$
resolved in our simulations.
For specific applications other weighting schemes may achieve even better results
and for $\bar{n}=3\times 10^{-4}({\rm Mpc/h})^3$
we can reduce $\sigma_n^2/P$ by up to a factor of 10 relative to uniform weighting.
These results open up new opportunities to extract cosmological information in galaxy surveys, such as 
the recently proposed multi-tracer approach to cancel sampling variance, and
may have important consequences for the planning of future redshift surveys.
Taking full advantage of these findings may require better understanding of galaxy formation process to 
develop accurate tracers of the halo mass. 

\end{abstract}


\maketitle

\setcounter{footnote}{0}


Galaxy clustering has been one of the leading methods to measure the clustering of dark matter 
in the past and with upcoming redshift 
surveys such as SDSS-III and JDEM/EUCLID this will continue to be the case in the future. 
Galaxies are easily observed and by measuring their redshift 
one can determine their three-dimensional distribution. This is currently the only large scale 
structure method that provides 3-dimensional information.
On large scales galaxies trace the dark matter 
up to a constant of proportionality called bias $b$, so the galaxy power spectrum 
can be directly related to the dark matter power spectrum shape, which contains a wealth of information 
such as the scale dependence of primordial fluctuations, signatures of massive neutrinos and matter density etc. 
In recent years 
the baryonic acoustic oscillations (BAO) feature in the power spectrum has been emphasized, 
which can be used as a standard ruler and in combination with 
cosmic microwave background anisotropies can provide a redshift distance test
\cite{2005ApJ...633..560E}. 

For the power spectrum measurement there are two sources of error:
one is the sampling (sometimes called cosmic) variance, the fact that each mode is a gaussian random realization 
and all the cosmological information lies in 
its variance, which cannot be well determined on the largest scales because the number of 
modes is finite. 
Second source of noise is the shot noise due to the discrete sampling 
of galaxies, 
$\sigma^2_n$, which under the standard assumptions of Poisson sampling equals the inverse of the number density $\bar{n}$.  
The total error on the power spectrum $P$ is $\sigma_P/P=(2/N)^{1/2}(1+\sigma_n^2/P)$, where $N$ is the number of modes
measured and scales linearly with the volume of the survey. 
While the above expression suggests there is not much benefit in reducing the shot noise to $\sigma_n^2/P \ll 1$ 
since sampling variance error remains, 
recent work suggests there are potential gains in that limit, since 
we may be able to reduce the damping of the BAO better \cite{2007ApJ...664..675E}. 

Recently a new multi-tracer method has been developed where by comparing two differently biased
tracers of the same structure one can extract 
cosmological information in a way that the sampling variance
error cancels out \cite{2009PhRvL.102b1302S}. There are several applications of this method, such as measuring the primordial 
non-gaussianity \cite{2009PhRvL.102b1302S}, redshift space distortion parameter $\beta$
\cite{2008arXiv0810.0323M}
or relation between the Hubble parameter and the angular diameter distance \cite{2008arXiv0810.0323M}.
In all these applications 
one can achieve significant gains in the error of the extracted cosmological parameters if
$\sigma_n^2/P \ll 1$.
Thus in all of these applications the galaxy shot noise relative to the power spectrum
is the key quantity that controls the ultimate 
level of cosmological precision one can achieve with galaxy surveys. 

The relation between the galaxy and the dark matter clustering can be understood with the halo model 
\cite{2000MNRAS.318.1144P,2000MNRAS.318..203S,2001ApJ...546...20S}, where all of the 
dark matter is divided into collapsed halos of varying mass. 
There are two contributions to the dark matter clustering:
first is the correlation between two separate halos, which is assumed to be proportional to the linear theory  
spectrum times the product of the two halo biases, while the second contribution is the one halo term which includes the clustering contributions from 
the individual halo itself. 
One obtains the dark matter power spectrum prediction by adding up the contributions from all the halos. 
Since galaxies are 
assumed to form inside the halos one can write analogous expressions for galaxy clustering power spectrum once one 
specifies the occupation distribution of galaxies as a function of halo mass. 

One consequence of the halo model is that the one halo term 
is dominated by the most massive halos and
reduces to white noise $k^0$ 
for very small wavemode amplitude $k \ll R^{-1}$, where $R$ is the  
size of the largest halos. 
For galaxies this is believed to be a valid description of the shot noise amplitude in the 
low $k$ limit. 
It distinguishes between the galaxy and the halo number density, but  
for a typical survey the fraction of halos with more than one galaxy in it is
small, 5-30\% \cite{2006MNRAS.368..715M}, and 
here we will ignore this distinction and assume for simplicity there is only one galaxy in each halo at its center.

For the dark matter, the 
nonlinear evolution of structure requires local mass and momentum  conservation and as a result the low $k$ limit of nonlinear contribution is 
predicted to scale as $k^4$ and not $k^0$ \cite{1980lssu.book.....P}. This is indeed seen in simulations \cite{2008PhRvD..77b3533C}, making this prediction 
of the halo model invalid. 
While this is often seen as a deficiency of the halo model, here we take it as an opportunity: if the 
dark matter has no white noise tail in the $k \rightarrow 0$ limit then in the context of the halo model 
where all the dark matter is in the halos and the halo size becomes irrelevant in $k \ll R^{-1}$ limit 
it should be possible  to achieve the same effect
with galaxies, if one can enforce the local mass and momentum conservation.
The most natural possibility is to weight the galaxies by the halo mass.

The purpose of this letter is to explore this idea with numerical simulations. 
We employ a suite of large N-body simulations using Gadget II code, which include four
$1024^3$ particles in a $(1.6\hGpc)^3$ box and one simulation with
$1536^3$ particles in a $(1.3\hGpc)^3$ box. 
The fiducial cosmological model 
has a scale invariant spectrum with amplitude 
$\sigma_8=0.81$, matter density $\Omega_m=0.28$ and Hubble parameter $H_0=70{\rm km/s/Mpc}$.
We ran Friends of Friends halo finder and kept all the halos with more than 20 particles, 
with the lowest halo mass of $6 \times 10^{12}\hMsun$ and $10^{12}\hMsun$, respectively. 

If a tracer has an overdensity $\delta_h$ with a bias $b_h$, then the relation to the dark matter 
overdensity $\delta_m$ in Fourier space
can be written as $\delta_h=b_h\delta_m +n$, where $n$ is shot noise with a power spectrum  
$\left< n^2 \right> =\sigma_n^2$ and we assume it is uncorrelated with the signal, ie $\left< \delta_m n \right>=0$ (the
operations should be taken separately on real and imaginary components of the Fourier modes). 
Thus we define $\sigma_n^2=\left< (\delta_h-b_h\delta_m)^2 \right>$ and bias is 
$b_h=(P_{hh}/P_{mm})^{1/2}=P_{hm}/P_{mm}$, where $P_{hh}=\left< \delta_h^2\right>-\sigma_n^2$, $P_{hm}=\left<\delta_m \delta_h \right>$
 and $P_{mm}=\left<\delta_m^2 \right>$.
This is equivalent to choosing $\sigma_n^2$ such that the cross correlation 
coefficient is unity, $r \equiv P_{hm}/(P_{hh} P_{mm})^{1/2}=1$. 
Thus our definition of the shot noise includes all sources of stochasticity between the halos and the 
dark matter, so it is the most conservative. 
This can be done as a function of $k$ and so allows for a possibility that noise is not white.
We do not assume a constant bias, although we find that for $k\ll 0.1{\rm h/Mpc}$  this is 
generally true. 
Another way to define the shot noise is through the power spectrum fluctuations,
$ \left< (\delta_h^2- P_{hh}-\sigma_n^2 )^2 \right> = (2/N)(P_{hh}^2+(\sigma_n^2)^2)$. We
find this definition in general has larger variance, but is on average in agreement with the definition above, which we will use in 
the following.

\begin{figure}
\centering
\includegraphics[scale=0.285,viewport=2.2cm 2.81cm 18.455cm 12.8cm,clip]{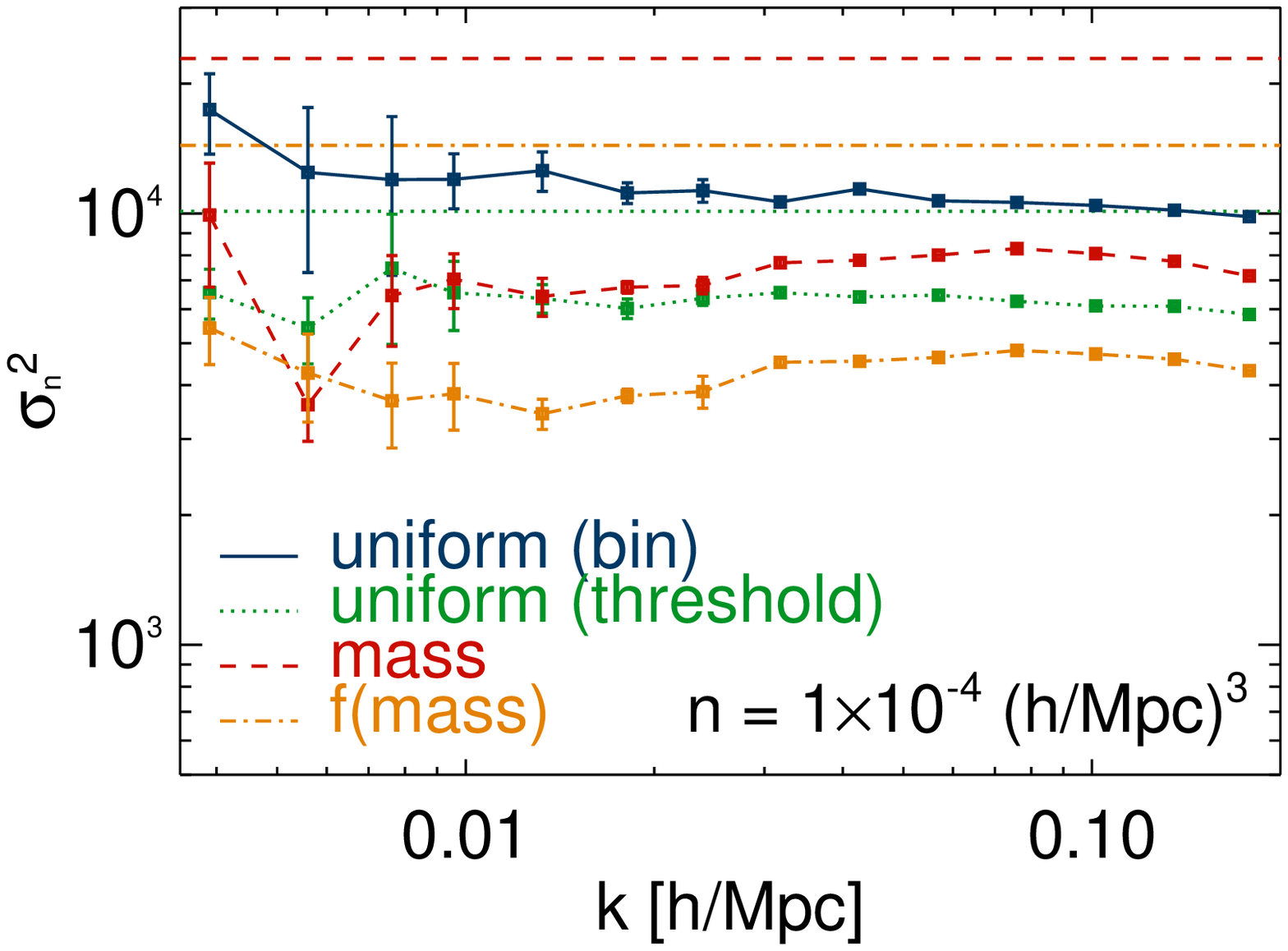}\includegraphics[scale=0.285,viewport=4.435cm 2.81cm 18.5cm 12.8cm,clip]{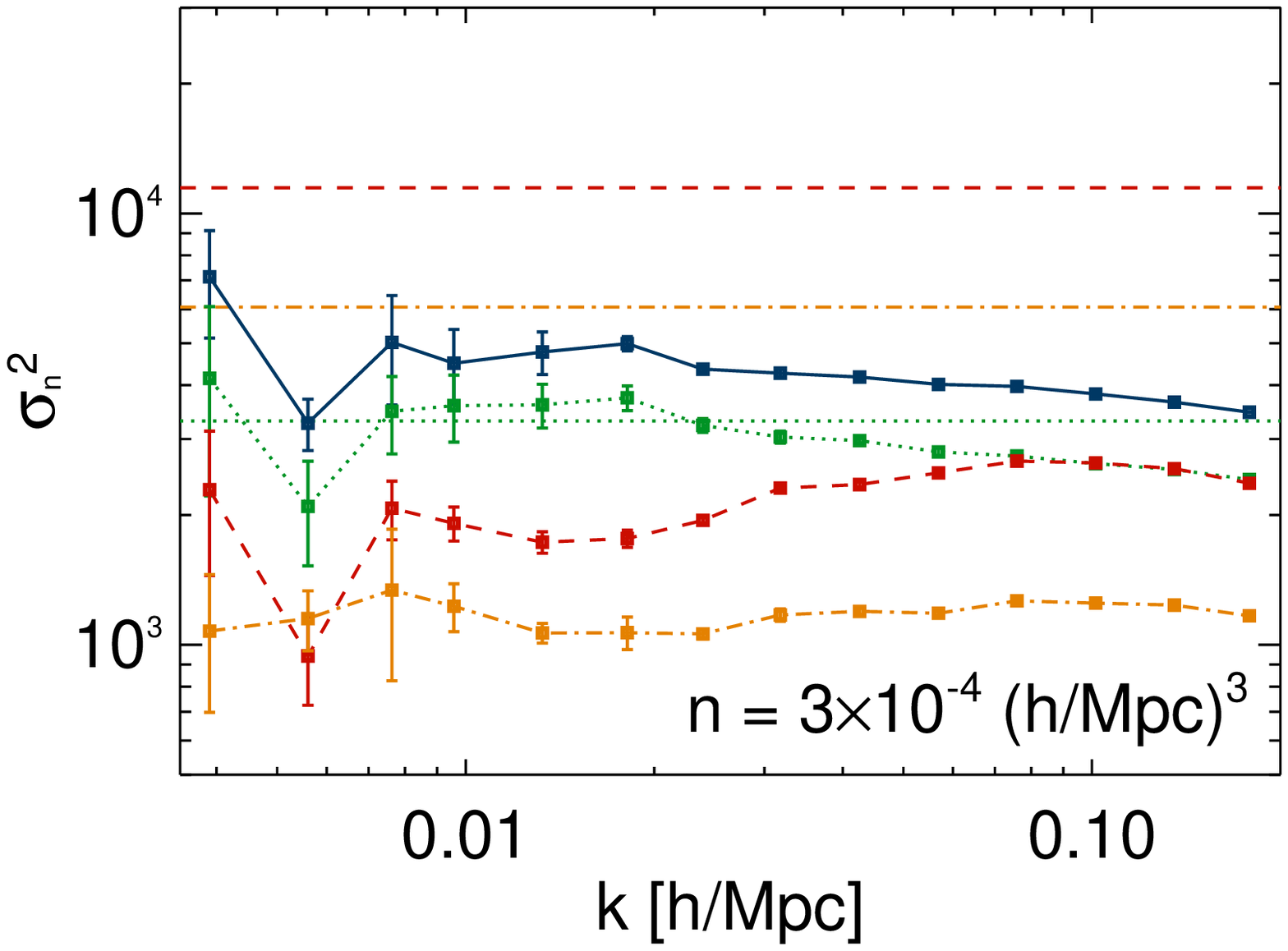}
\includegraphics[scale=0.285,viewport=2.2cm 0.6cm 18.455cm 12.6cm,clip]{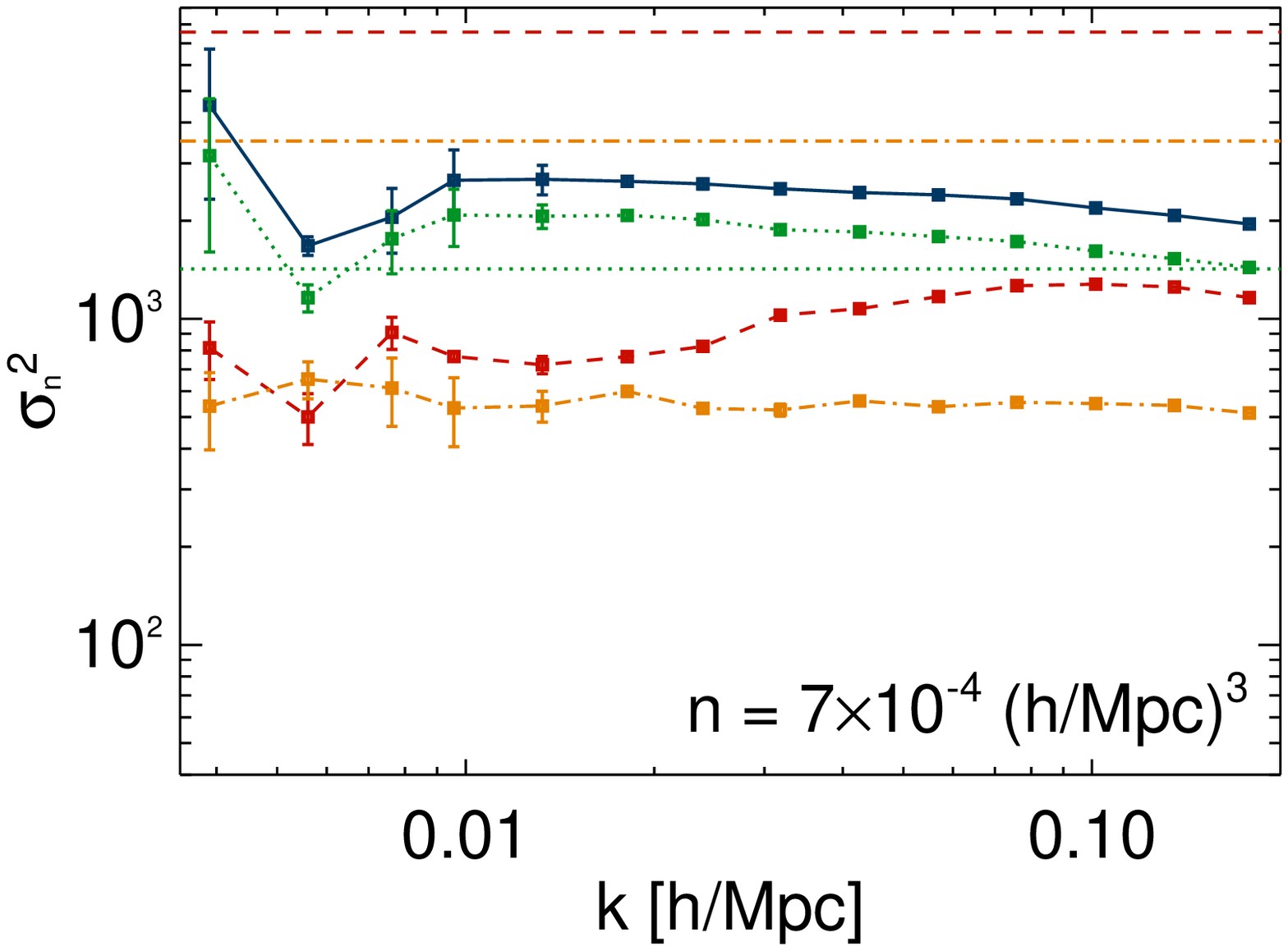}\includegraphics[scale=0.285,viewport=4.435cm 0.6cm 18.5cm 12.6cm,clip]{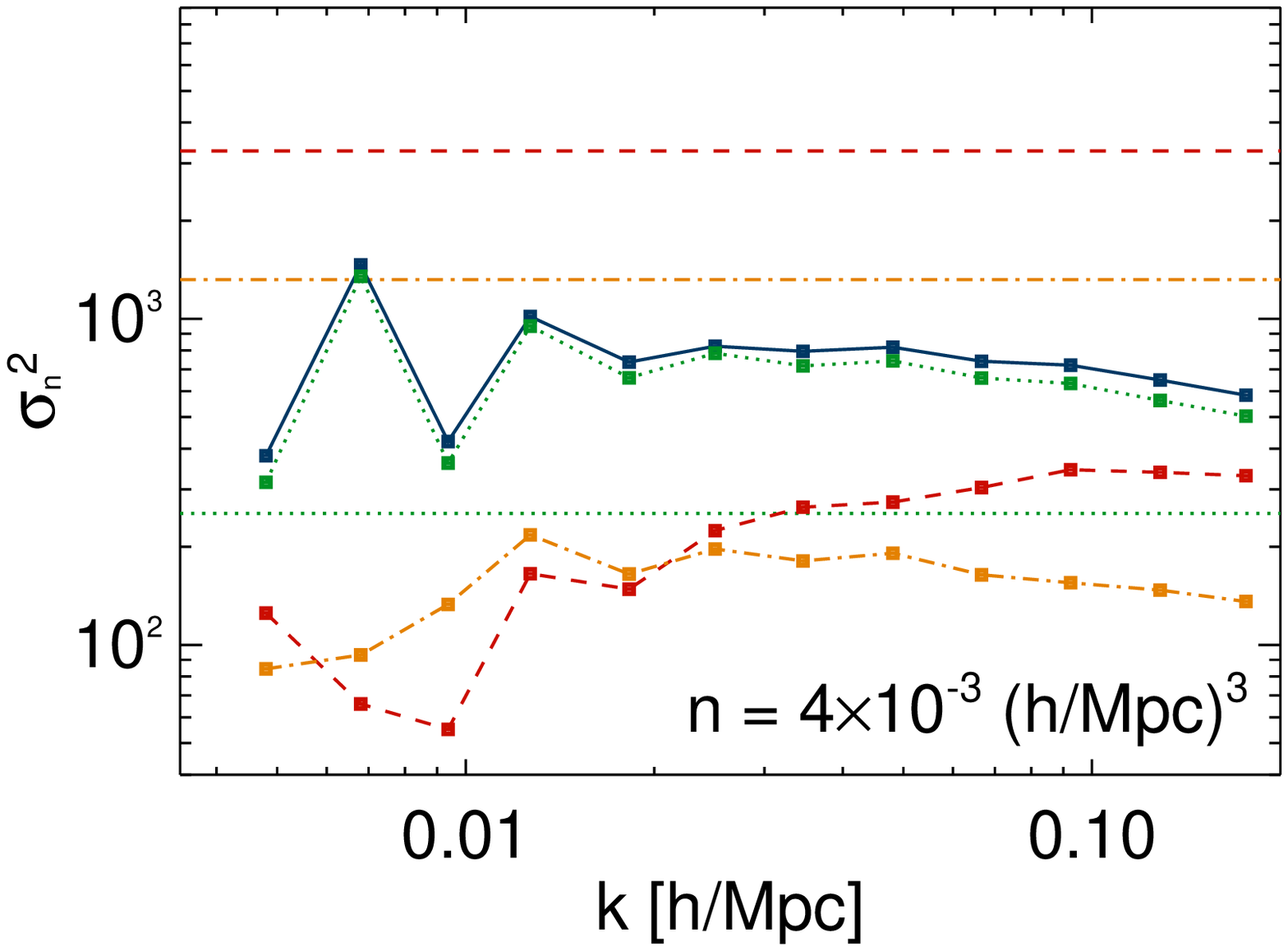}
\caption{Shot noise power spectrum $\sigma_n^2$ 
measured in simulations for uniform weighting of halos in a mass bin and mass 
threshold, mass weighting and $f(M)=M/(1+(M/10^{14}\hMsun)^{0.5}$ weighting, for 
several different abundances, corresponding at $z=0$ to mass thresholds of $4\times 10^{13}\hMsun$, $1.4 \times 10^{13}\hMsun/h$, 
$6 \times 10^{12}\hMsun$ and $10^{12}\hMsun/h$, from the lowest to the highest abundance, respectively. 
Straight lines (same color/line style) are the expected shot noise $\sigma_e^2$ for 
each of the weightings (equal for the mass bin and mass threshold with uniform weighting).}
\label{fig1}
\end{figure}

We begin by first investigating the shot noise 
when each halo has equal weight. The simplest case is that of 
a bin in halo mass, for which we remove the top 10\% of the most massive halos in a simulation and take the 
remaining ones to match a given abundance.
As shown in figure \ref{fig1} the 
prediction $\sigma_n^2=\bar{n}^{-1}$ is satisified for $\bar{n}=10^{-4}({\rm h/Mpc})^3$, but is lower than simulations 
at higher abundances, by a factor of 3 for our highest number density of $\bar{n}=4 \times 10^{-3}({\rm h/Mpc})^3$. 
Second possibility is that of a mass threshold, where all of the halos above certain minimum cutoff are 
populated. This gives a bit lower shot noise than mass bins for the same abundance. Overall, we find that 
weighting the galaxies uniformly leads to a shot noise power 
that can be larger than the usually assumed $\bar{n}^{-1}$, so the standard error 
analysis may be overly optimistic and shot noise should 
be a free parameter determined from the data itself.

Next we investigate the shot noise for non-uniform halo dependent weighting $w_i$ for the same mass threshold sample. 
We compare the simulations to the expectation $\sigma^2_e=V \sum_i w_i^2/(\sum_i w_i)^2$, where 
$V$ is the volume and the sum is over all the halos. 
At a given number density this expression is minimized for uniform weighting (where it equals $\bar{n}^{-1}$), 
so non-uniform weighting generally increases the expected shot noise. 
As argued above $w_i=M_i$, where $M_i$ is the halo mass, 
is the natural implementation of the idea to enforce mass and momentum conservation for the halos. 
The results are shown in figure \ref{fig1}.
We see that 
the predicted and measured shot noise amplitudes differ significantly and the difference 
reaches a factor of 10-30  
at the highest abundance in our simulations, $\bar{n}=4 \times 10^{-3}({\rm h/Mpc})^3$.
This demonstrates that this is not a simple Poisson 
sampling of the field and that mass and momentum conservation work to suppress
the shot noise relative to expectations. 

Other weightings may also improve the results relative to naive expectations and may work even better 
for specific applications. 
For example, weighting by $f(M)=M/(1+(M/10^{14}\hMsun)^{0.5})$,
shown in figure \ref{fig1}, improves upon the mass 
weighting.
This weighting equals the halo mass weighting over the mass range of $M<10^{14}\hMsun$, 
while giving a lower weight to the higher mass halos relative to the mass weighting.  
Weighting by the halo mass gives a very large weight to the most massive halos and this non-uniform weighting leads 
to a significant increase in the naive shot noise prediction $\sigma_e^2$ relative to the number density of halos. 
Therefore, if the conservation of mass and momentum is not perfect for the most massive halos
the residual shot noise may still be large, which may explain why downweighting high mass halos
may work better.  On the other hand, simply eliminating the halos above $10^{14}\hMsun$ while preserving mass
weighting below that mass completely erased any advantages. 
We also tried weighting by the halo bias $b$, which was argued to minimize $\sigma^2/P$
\cite{2004MNRAS.347..645P}, and found no improvements relative to uniform weighting, as expected
since it is close to uniform weighting for most of the halos and therefore 
does not implement the mass and momentum conservation efficiently. 
It is possible that one may be able to further improve 
the signal to noise by optimizing the weights, but 
the optimization will depend on the specific application one has in mind 
(e.g. non-gaussianity, redshift space distortions, BAO etc.) and is beyond the scope of this paper.  
 
\begin{figure}
\centering
\includegraphics[scale=0.285,viewport=2cm 2.81cm 18.455cm 12.8cm,clip]{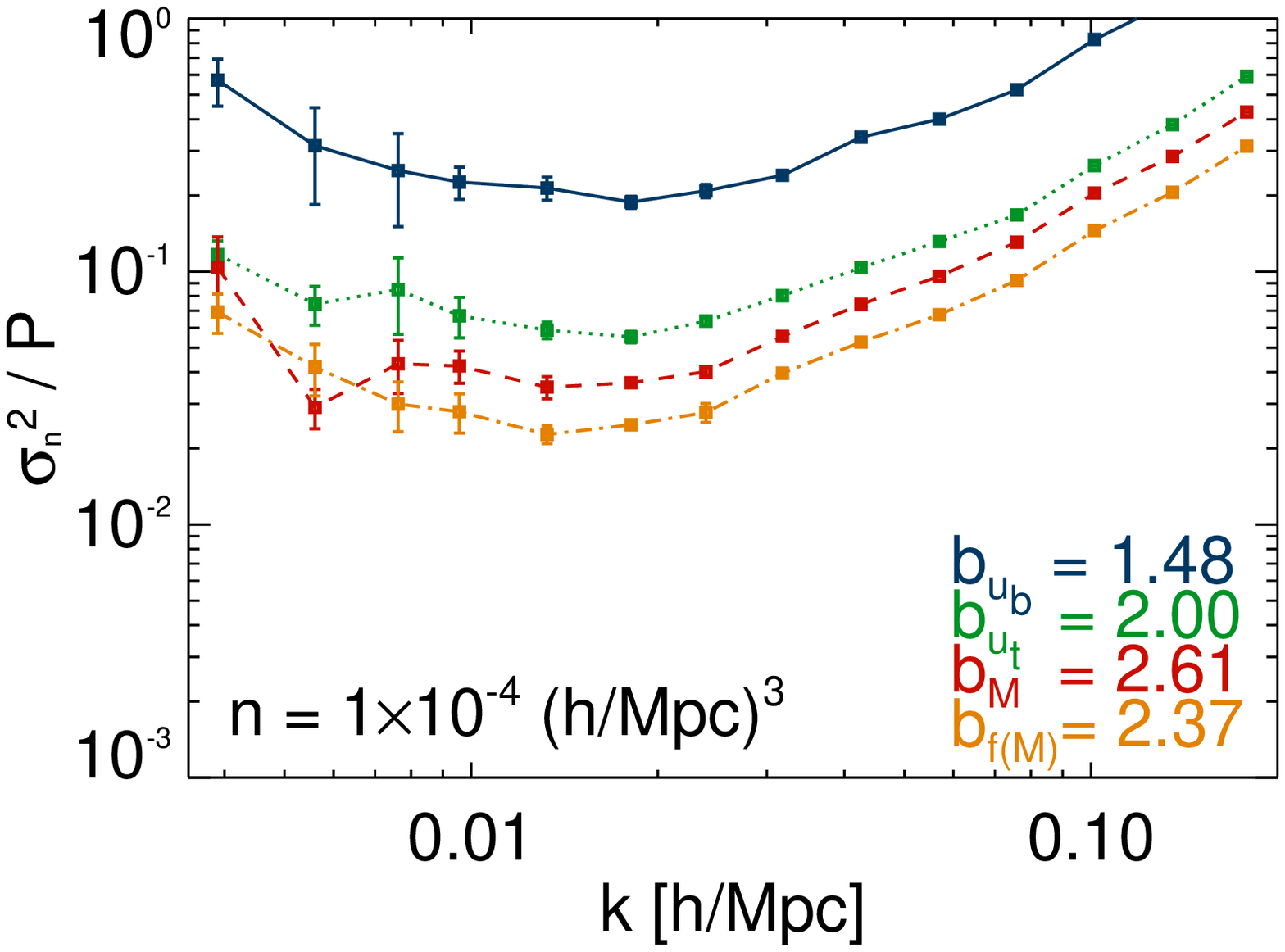}\includegraphics[scale=0.285,viewport=4.435cm 2.81cm 18.5cm 12.8cm,clip]{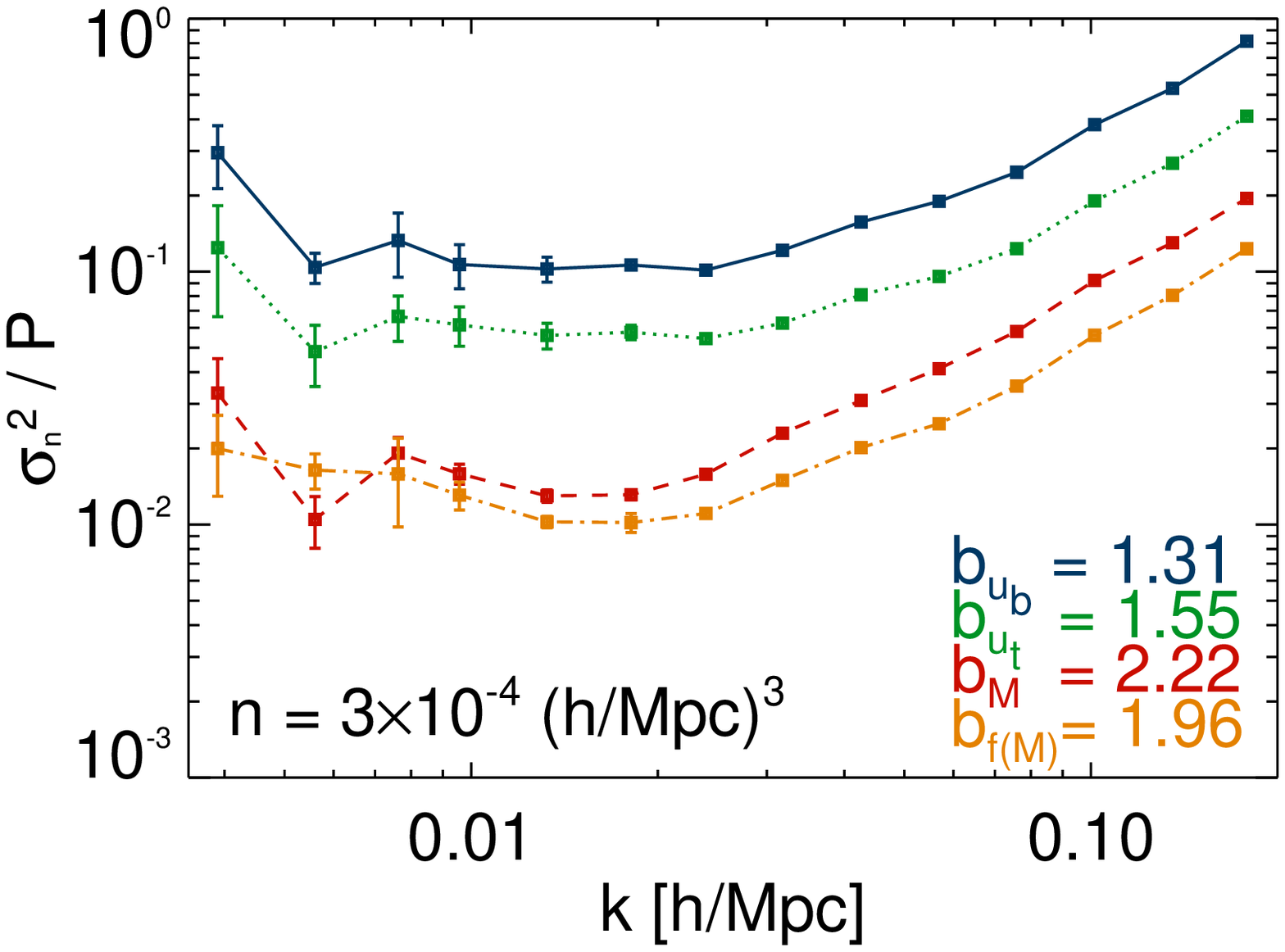}
\includegraphics[scale=0.285,viewport=2cm 0.6cm 18.455cm 12.6cm,clip]{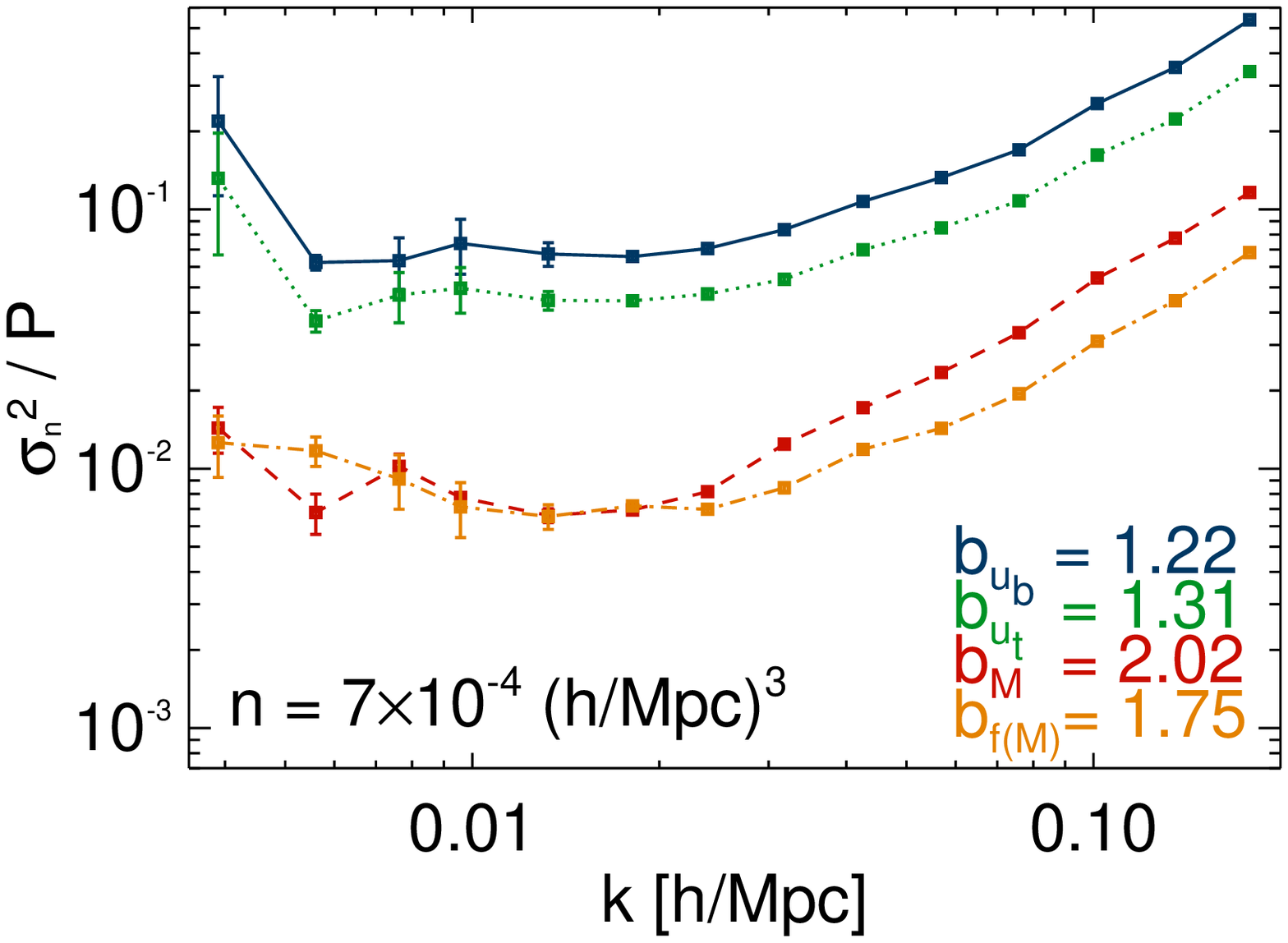}\includegraphics[scale=0.285,viewport=4.435cm 0.6cm 18.5cm 12.6cm,clip]{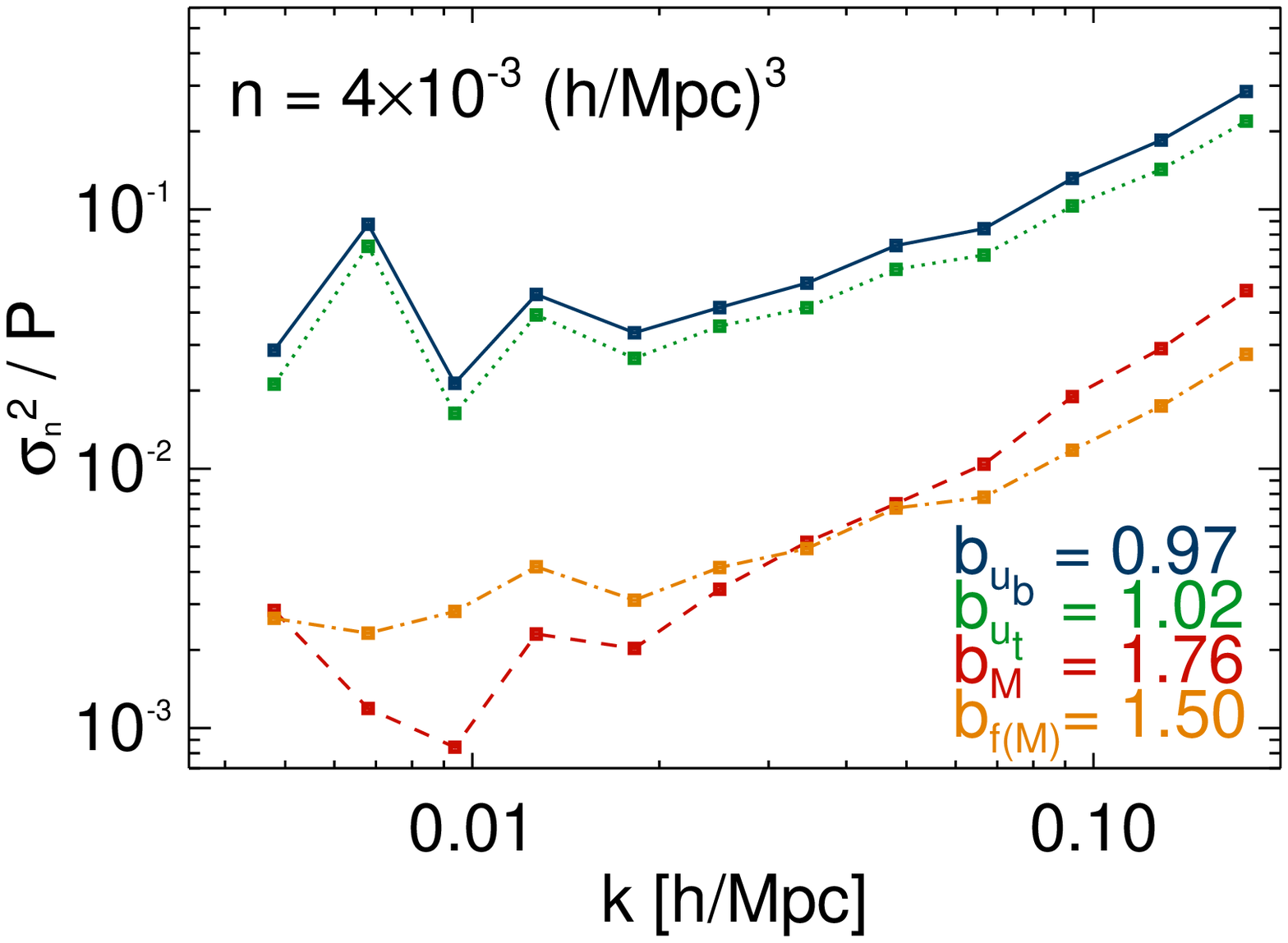}
\caption{Same as figure \ref{fig1}, but for $\sigma_n^2/P$. Also shown are the bias values for the 
different cases. }
\label{fig2}
\end{figure}

For actual applications we want to minimize $\sigma_n^2/P$. 
Figure \ref{fig2} shows the results for  
the same cases as in figure \ref{fig1}. We see there are significant improvements in $\sigma_n^2/P$ relative to the uniform
weighting and that mass and modified mass give comparable results, with improvements in excess of 10 possible relative to the 
uniform weighting. 
While these results are all at $z=0$ where we have the highest density of halos, we also computed them 
at higher redshifts. At $z=0.5$ and $\bar{n}=3\times 10^{-4}({\rm h/Mpc})^3$, target density for SDSS-III,
we find a factor of 3-10 improvement at BAO scale
in mass weighting relative to the uniform, comparable to $z=0$ case at the same number density. 
This means that the achievable error on cosmological parameters from BAO
can be improved significantly for the same number of objects measured. Alternatively, a significantly lower 
number of objects may be needed to achieve the same precision and 
one can reduce the target number density by nearly a factor of 3. Note that SDSS-III plan is to oversample
the galaxies at the BAO scale to use reconstruction to reduce the damping of BAO, which can be done better 
if the shot noise is lower.  
It is also possible that imposing the local mass and momentum conservation will minimize systematic 
shifts in the BAO position relative to the dark matter that may otherwise be problematic \cite{2008PhRvD..77d3525S}, 
but we leave this investigation for the future.

\begin{figure}
\centering
\includegraphics[scale=0.285,viewport=2.2cm 2.81cm 18.455cm 12.8cm,clip]{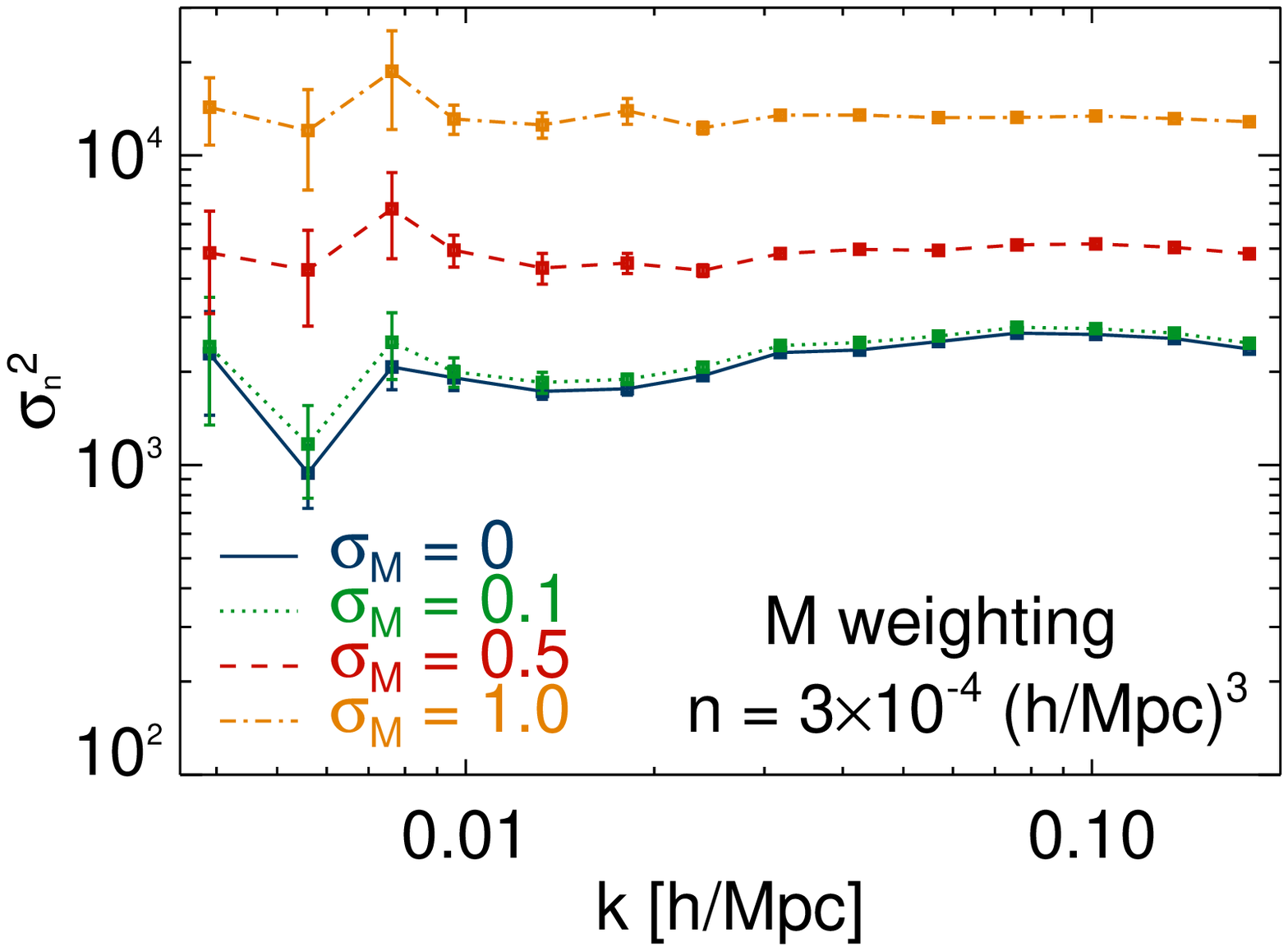}\includegraphics[scale=0.285,viewport=4.435cm 2.81cm 18.5cm 12.8cm,clip]{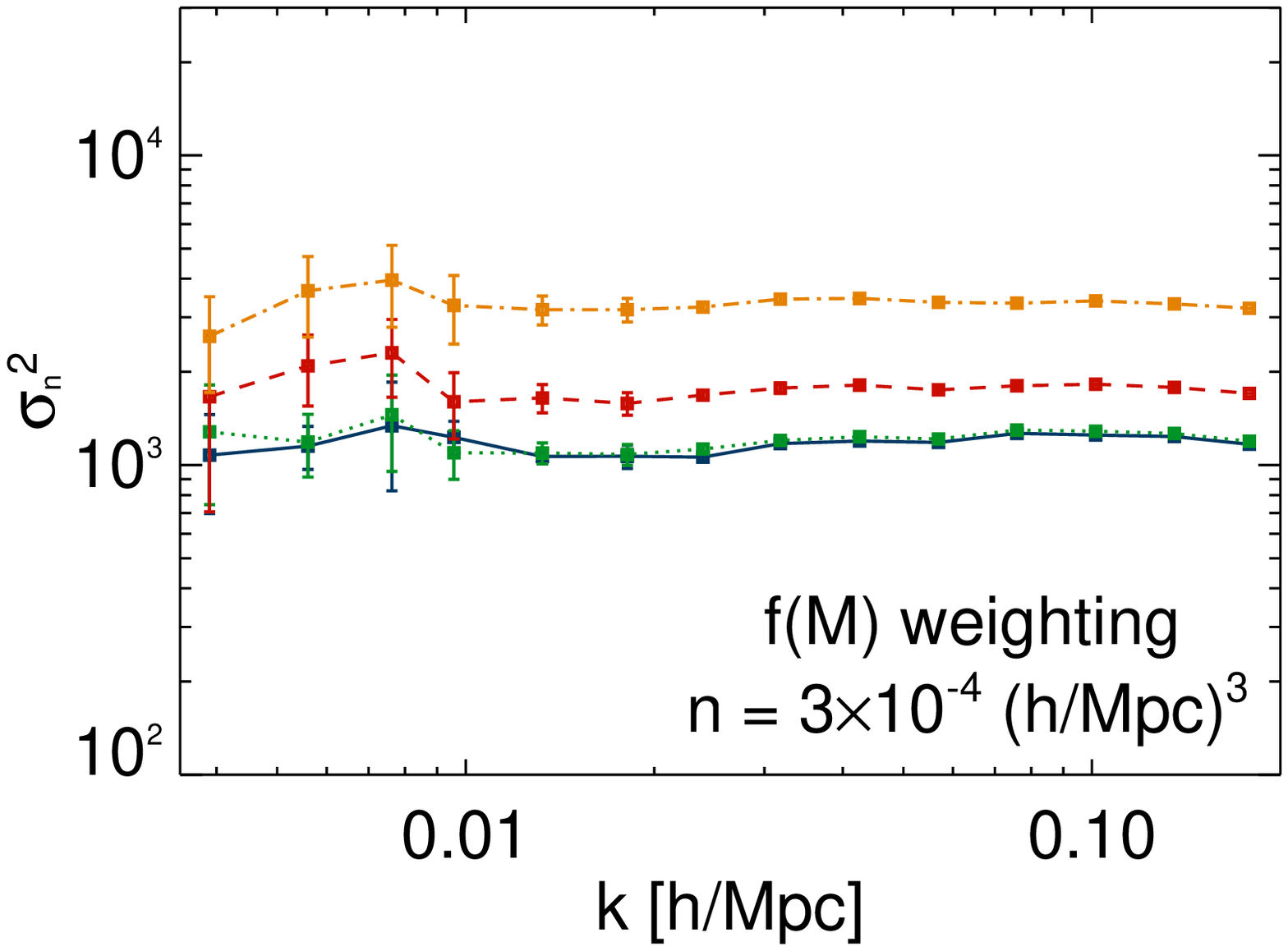}
\includegraphics[scale=0.285,viewport=2.2cm 0.6cm 18.455cm 12.6cm,clip]{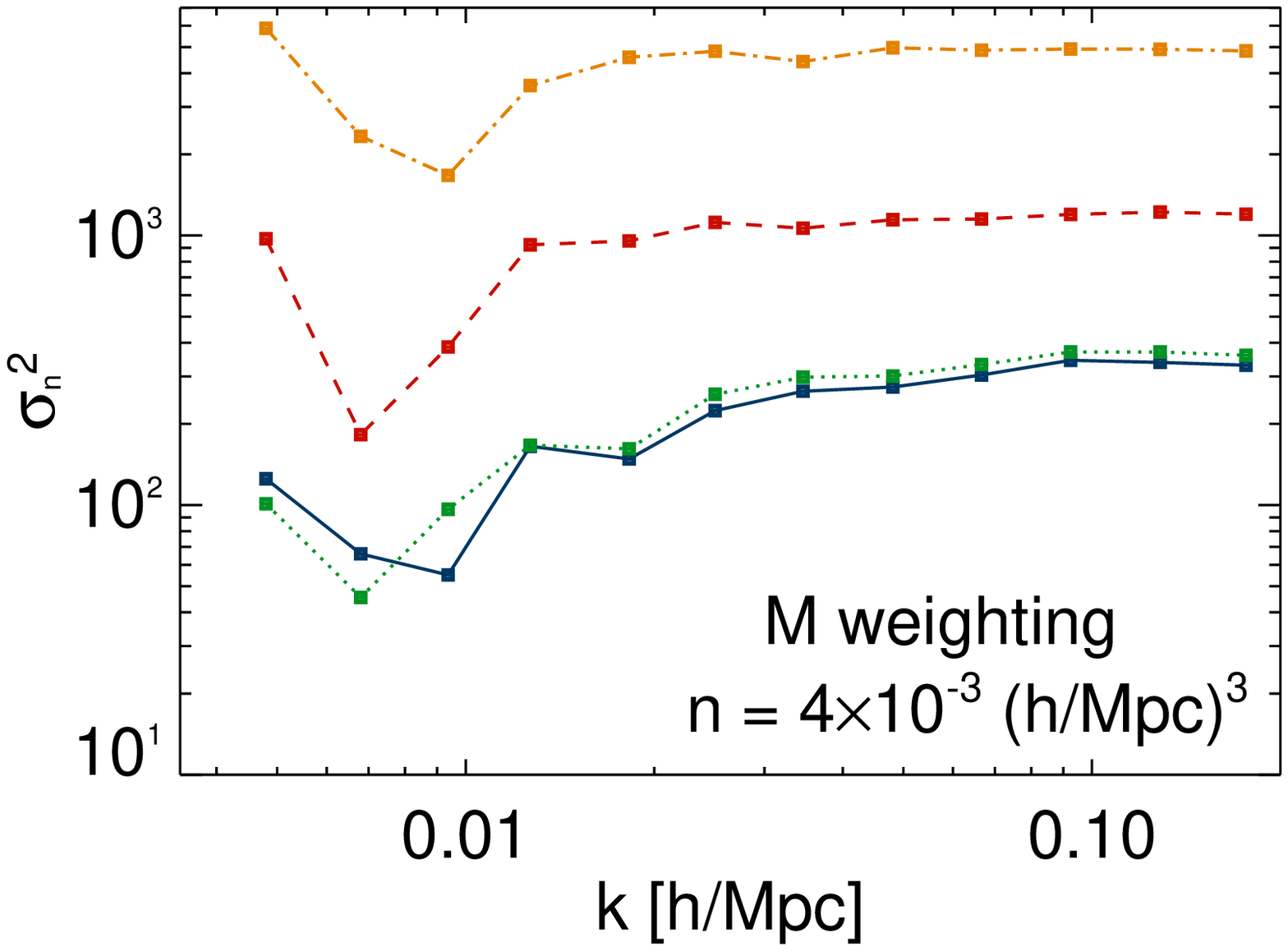}\includegraphics[scale=0.285,viewport=4.435cm 0.6cm 18.5cm 12.6cm,clip]{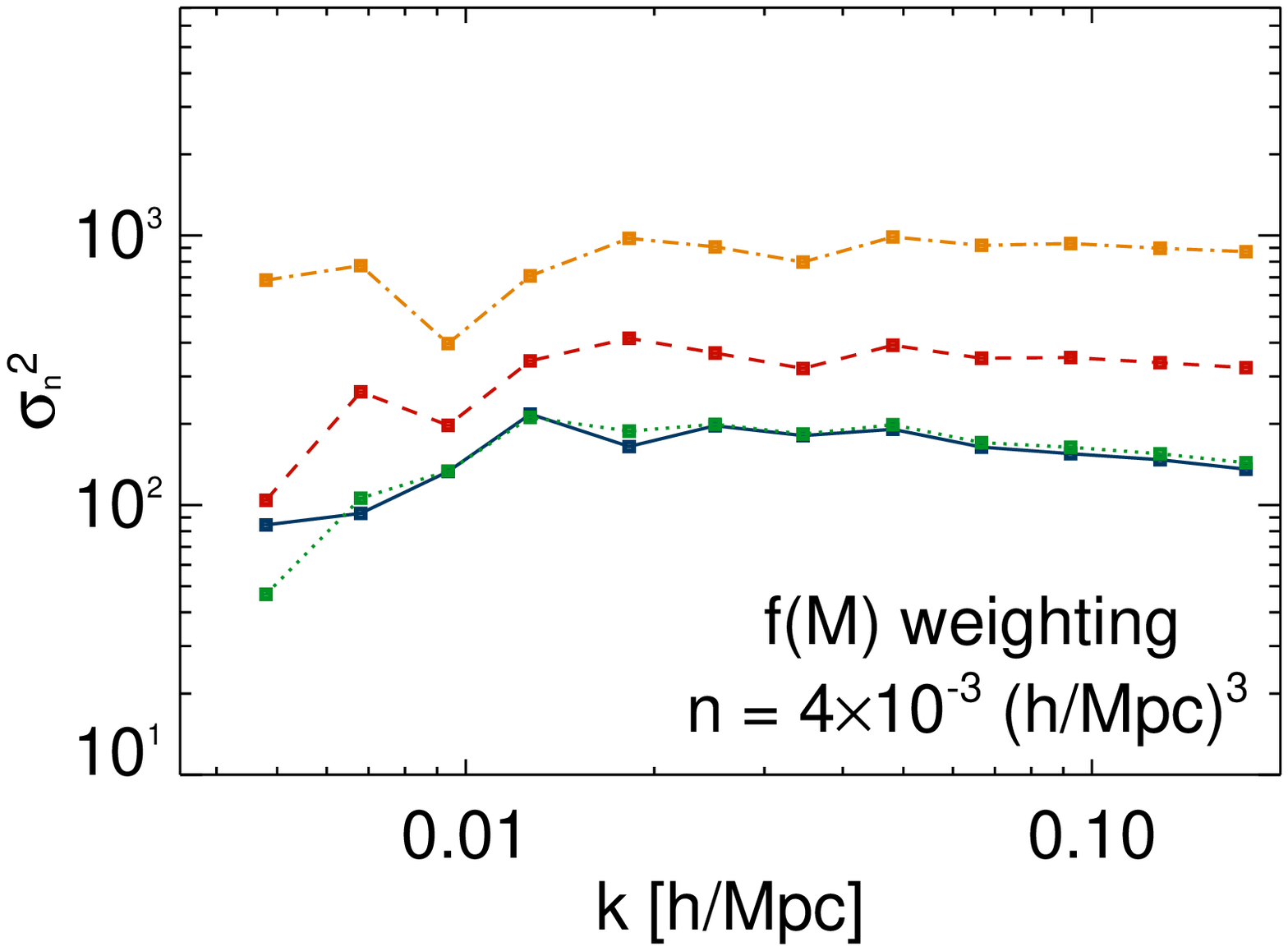}

\caption{Effects of log-normal scatter $\sigma$ in halo mass observable on the shot noise $\sigma_n^2$ for mass and $f(M)=M/(1+(M/10^{14}\hMsun)^{0.5})$
weights, for $\bar{n}=3\times 10^{-4}({\rm h/Mpc})^3$ and $\bar{n}=4\times 10^{-3}({\rm h/Mpc})^3$. Scatter hardly affects the bias, so 
the relative effects of scatter are the same for $\sigma_n^2/P$ and we do not show them here.}
\label{fig3}
\end{figure}

So far we ignored the real world complications such as the imprecise knowledge of the halo mass. 
To investigate this we add a 
log-normal scatter with rms variance $\sigma$ to each halo mass and recompute the analysis. Fig \ref{fig3} shows the results for 
mass and modified mass $f(M)$ weighting: for the latter we see that scatter of 50\% in mass
increases $\sigma_n^2/P$ by about 50\% for lower abundance and a factor of 2 for higher abundance. Since this is a realistic 
scatter for optically selected clusters \cite{2008arXiv0809.2794R} there is 
thus realistic possibility that we can apply such analysis to the real data and 
achieve these gains. In practical applications one would try to identify the best halo mass tracer as a function of halo mass, for 
example central galaxy luminosity in the galactic halos and richness or total luminosity for the cluster halos. 
In order to minimize the scatter one must understand the relation between the galaxy observables and the underlying halos, 
so progress in galaxy formation studies will be needed to maximize the gains. 
We find that for the mass weighting scatter has a larger effect, such that for $\sigma=0.5$
the degradation in $\sigma_n^2/P$ is a factor of 2-3. Once the scatter becomes too large there is no longer any 
local mass and momentum conservation and we find that for $\sigma=1$ the shot noise is worse than for uniform weighting. 
Another potential complication is the effect of redshift space distortions, 
since the observed radial distance is a sum of the true radial distance 
and peculiar velocity (divided by the Hubble parameter).
We find a modest (50\%) increase in $\sigma_n^2/P$, where $P$
in redshift space
is the spherically averaged (i.e. monopole) power spectrum. 
Since redshift space contains much more information than just the monopole 
it is possible that one may be able to use the additional information to 
reduce this degradation and we leave this for a future investigation. 

These results are particularly relevant for the multi-tracer methods where the data are analyzed in terms of ratios of 
different tracers and 
for which the sampling variance error cancels, such as 
those recently proposed for non-gaussianity \cite{2009PhRvL.102b1302S}, redshift space distortions and Hubble versus angular
distance relation \cite{2008arXiv0810.0323M}. For these there is no lower limit on the
achievable error decreases as long as $\sigma_n^2/P$ decreases and 
the method proposed here could lead to a significant reduction of 
errors relative to previous expectations. We see from figure \ref{fig2} that 
for mass weighting at $4 \times 10^{-3}({\rm h/Mpc})^3$ $\sigma_n^2/P \sim 10^{-3}$ on large scales, so this 
could give a signal to noise of 30 for a single mode, compared to 0.7 for the single tracer method, equivalent to 3 orders of 
magnitude reduction in volume needed to reach the same precision.
Note that this is not unreachable, since the existing SDSS survey achieves $\bar{n} \sim 10^{-2} ({\rm h/Mpc})^3$ for the 
redshift survey of the main sample. 

Equally impressive improvements may be possible for future redshift surveys such as  
JDEM/EUCLID or BigBOSS, which are expected to operate at redshifts up to $z \sim 2$. 
Their target number density could be as high as $\bar{n}\sim 10^{-3}({\rm h/Mpc})^3$ or higher, and the method proposed here 
could lead to a dramatic reduction of errors or, equivalently, to a 
several-fold reduction in the number of measured redshifts required to reach the target precision, with 
potentially important implications for the design of these missions. 
The weights can be further 
optimized for specific applications, specially for the multi-tracer methods that cancel out the sampling variance error.
This approach holds the promise to become the most accurate method to extract both the primordial non-gaussianity 
and the dark energy equation of state and its full promise should 
be explored further with more realistic simulations. In parallel we should 
develop better our understanding of galaxy formation to relate the galaxy observables to the underlying halo mass 
with as little scatter as possible. 

We thank V. Springel for making the Gadget-2 code available to us and P. McDonald, D. Eisenstein and R. Smith for 
useful comments. 
This work is supported by the
Packard Foundation, the
Swiss National Foundation
under contract 200021-116696/1 and WCU grant R32-2008-000-10130-0.

\bibliography{cosmo,cosmo_preprints}
\end{document}